\documentclass[reprint,superscriptaddress,aip,amsmath,amssymb,floatfix]{revtex4-1}

\pdfoutput=1

\usepackage{graphicx}
\usepackage{bm}
\usepackage{epstopdf}
\usepackage[pdftex,breaklinks=true,bookmarksopen=true,bookmarksopenlevel=3,bookmarksnumbered=true,colorlinks=true,urlcolor= magenta,citecolor=blue,linkcolor=blue]{hyperref}

\begin{document}

\title{All-optical Compton gamma-ray source}

\author{K. Ta Phuoc}
\author{S. Corde}
\author{C. Thaury}
\author{V. Malka}
\author{A. Tafzi}
\author{J. P. Goddet}
\affiliation{Laboratoire d'Optique Appliqu\'ee, ENSTA ParisTech - CNRS UMR7639 - \'Ecole Polytechnique, 828 Boulevard des Mar\'echaux, 91762 Palaiseau, France}
\author{R. C. Shah}
\affiliation{Los Alamos National Laboratory, Los Alamos, New Mexico, USA}
\author{S. Sebban}
\author{A. Rousse}
\affiliation{Laboratoire d'Optique Appliqu\'ee, ENSTA ParisTech - CNRS UMR7639 - \'Ecole Polytechnique, 828 Boulevard des Mar\'echaux, 91762 Palaiseau, France}


\maketitle

\textbf{
One of the major goals for laser-plasma accelerators \cite{Esarey} is the realization of compact sources of femtosecond X-rays \cite{Rousse, Kneip, Dino}. In particular, using the modest electron energies obtained with existing laser systems, Compton scattering a photon beam off a relativistic electron bunch has been proposed as a source of high-energy and high-brightness photons. However, laser-plasma based approaches to Compton scattering have not, to-date, produced X-rays above 1 keV. Here, we present a simple and compact scheme for a Compton source based on the combination of a laser-plasma accelerator and a plasma mirror. This approach is used to produce a broadband spectrum of X-rays extending up to hundreds of keV and with 10,000-fold increase in brightness over Compton X-ray sources based on conventional accelerators \cite{Schoenlein,Albert}.
We anticipate that the technique will lead to compact, high-repetition-rate sources of ultrafast (femtosecond), tunable (X- through gamma-ray), and low-divergence ($\sim1^\circ$) photons from source sizes on the order of a micrometre.
}
\begin{figure}[b]
\includegraphics[width=\linewidth]{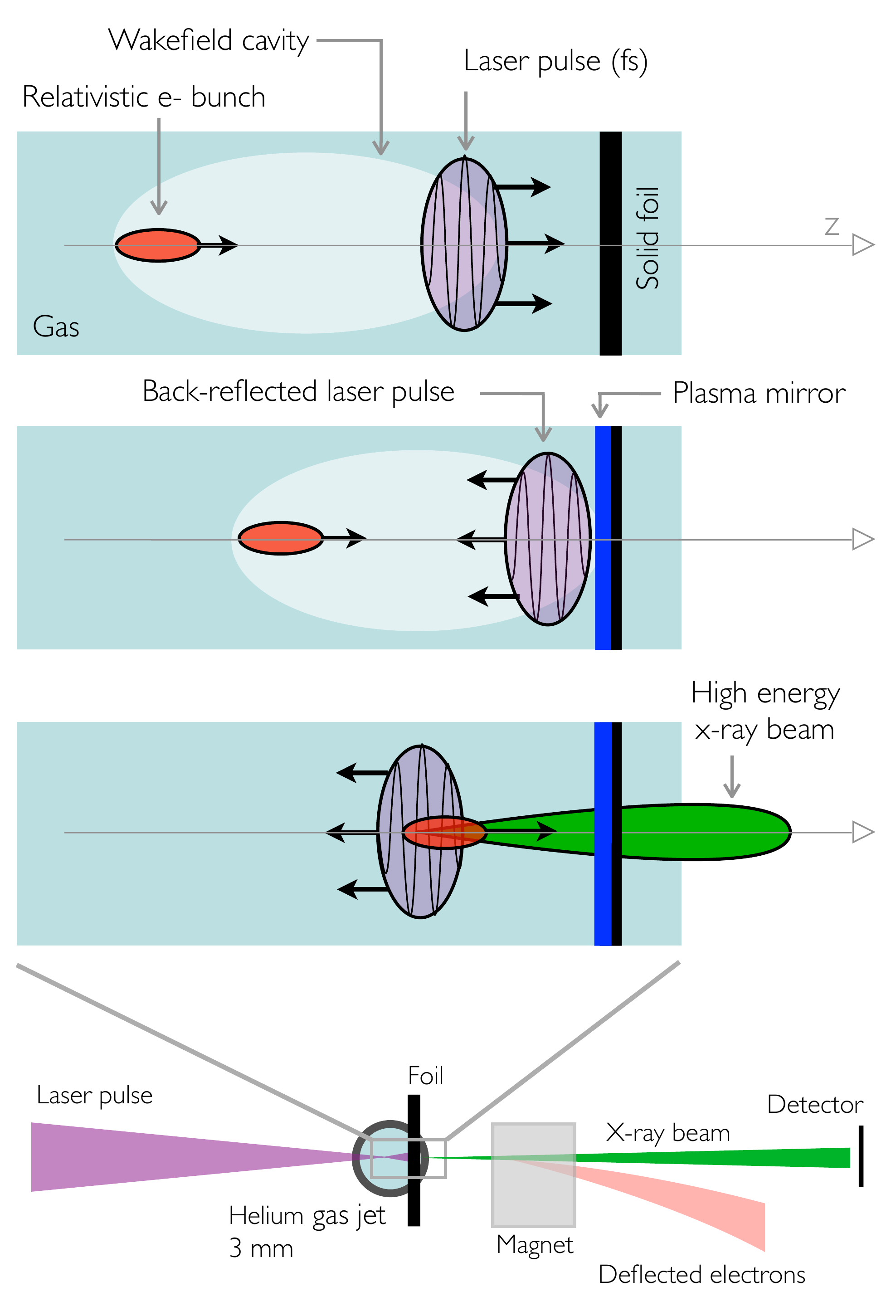}
\caption{\textbf{Principle of the Compton backscattering source.} An intense laser pulse drives a plasma accelerator and is reflected by a plasma mirror. At the collision between the back-reflected laser pulse and the relativistic electron bunch, a femtosecond X-ray pulse is emitted in the forward direction. Lower part of the figure represents the experimental set-up. A 30 TW (30 fs) laser pulse is focused onto a supersonic, 3-mm-diameter, helium gas jet. The radiation is measured using a phosphor screen imaged with a CCD camera or with an imaging plate.}
\label{fig1}
\end{figure}

A Compton scattering source operates by colliding a relativistic electron bunch and an intense laser pulse. The electrons traveling in the electromagnetic field oscillate and emit synchrotron-like radiation, commonly referred to as Compton scattering radiation. The scheme provides a double Doppler upshift of incident photon energy by relativistic electrons \cite{Hartemann1}. For an entirely optical realization of this scheme, two intense laser pulses are required \cite{Catravas,Hartemann2,Schwoerer}, the first to create a plasma accelerator and the second to scatter off the accelerated relativistic electrons. While attractive in its potential, this scheme has never yet produced radiation above a few keV \cite{Schwoerer}. Here, we demonstrate high-energy femtosecond X-ray radiation from Compton backscattering (hundreds of keV range) in a purely optical scheme which is both simple and robust. 
As shown in Fig. \ref{fig1}, an intense  femtosecond laser pulse focused into a millimeter-scale gas jet drives a wake-field cavity in which electrons are trapped and accelerated. This is the laser-plasma accelerator. A foil is positioned close to the exit of the gas jet and at nearly normal incidence with respect to the laser and electron beam axis ${\hat z}$. The foil is ionized by the rising edge of the laser pulse, resulting in a plasma mirror \cite{Kapteyn} that efficiently reflects the laser pulse (the reflectivity is expected to be larger than 70\% for $a_0\gtrsim 0.04$ \cite{Doumy}). This approach provides an inherent overlap in time and space of the back-reflected laser and the relativistic electrons for Compton backscattering.

\begin{figure}[t]
\includegraphics[width=\linewidth]{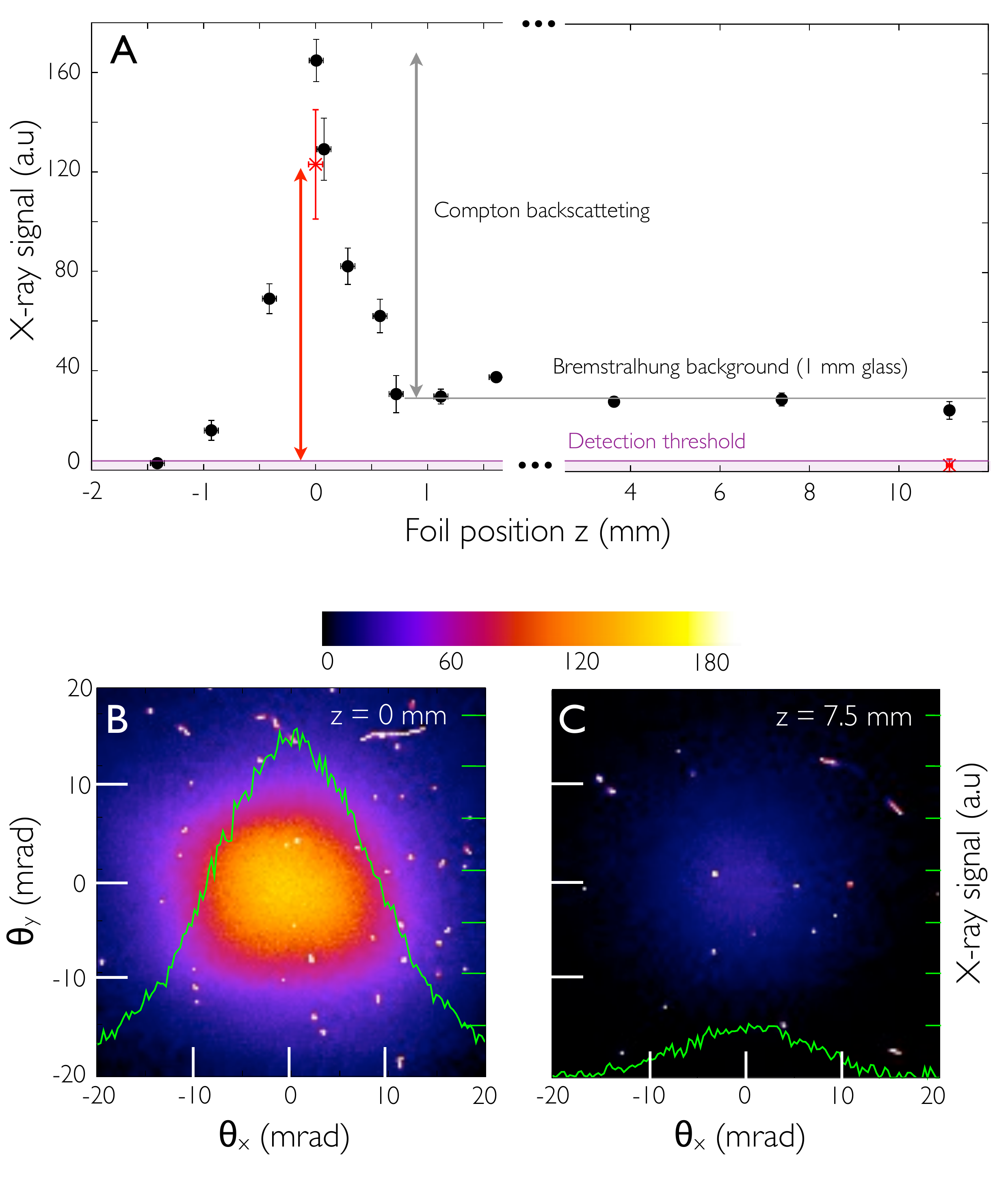}
\caption{\textbf{Evolution of the X-ray signal as a function of the foil position.} {\bf (A)} $z$ is the foil position along the laser propagation axis (the $z$-axis is oriented in the direction of the laser propagation). Each data point corresponds to an average over five shots, and the error bar give the standard error of the mean. The foil is 1 mm of glass for data in black circles and  300 $\mu$m CH for data indicated by red crosses. {\bf (B)} Beam profile observed in a single shot for $z=0$ mm. {\bf (C)} Beam profile observed in a single shot for $z=7.5$ mm. Radiation below $\sim 10$ keV was blocked with filters.}
\label{fig2}
\end{figure}

The experimental set-up is presented in the lower part of Fig. \ref{fig1}. The relativistic laser pulse is focused into a 3 mm gas jet to produce an electron beam (see Methods). For the sake of simplicity and reproducibility, the laser-plasma accelerator was set to produce electrons in the forced laser wake-field regime \cite{Malka}, with a broadband spectrum extending up to an energy $E \sim 100$ MeV (corresponding to a Lorentz factor $\gamma = E[\textrm{MeV}]/0.511$). The measured charge in an electron bunch was about 120 pC. The foil used was either 1 mm of glass or 300 microns of CH. Supplementary Fig. S1 presents electron spectra obtained with (i) the foil in the gas jet, (ii) foil with a hole in it corresponding to the laser propagation axis, (iii) foil outside the gas jet (more than one centimeter after the gas jet). It shows that the presence of the foil or interaction with the foil do not significantly improve or degrade the properties of the electron beam.
 
In this relativistic interaction, radiative processes other than Compton backscattering can produce X-ray radiation. Careful measurements were performed to identify, evaluate and minimize their contribution to the overall emission. In our parameter regime, betatron radiation is emitted in the few keV range and about $10^8-10^9$ photons are produced \cite{Rousse}. This radiation was measured and blocked by filters (glass and aluminum). Relativistic Bremsstrahlung radiation can be produced when the electron bunch crosses the foil \cite{Jackson}. Bremsstrahlung radiation is dependent on the material and thickness of the foil and was minimized by using a thin low-Z material foil for the plasma mirror. 

\begin{figure}[b]
\includegraphics[width=\linewidth]{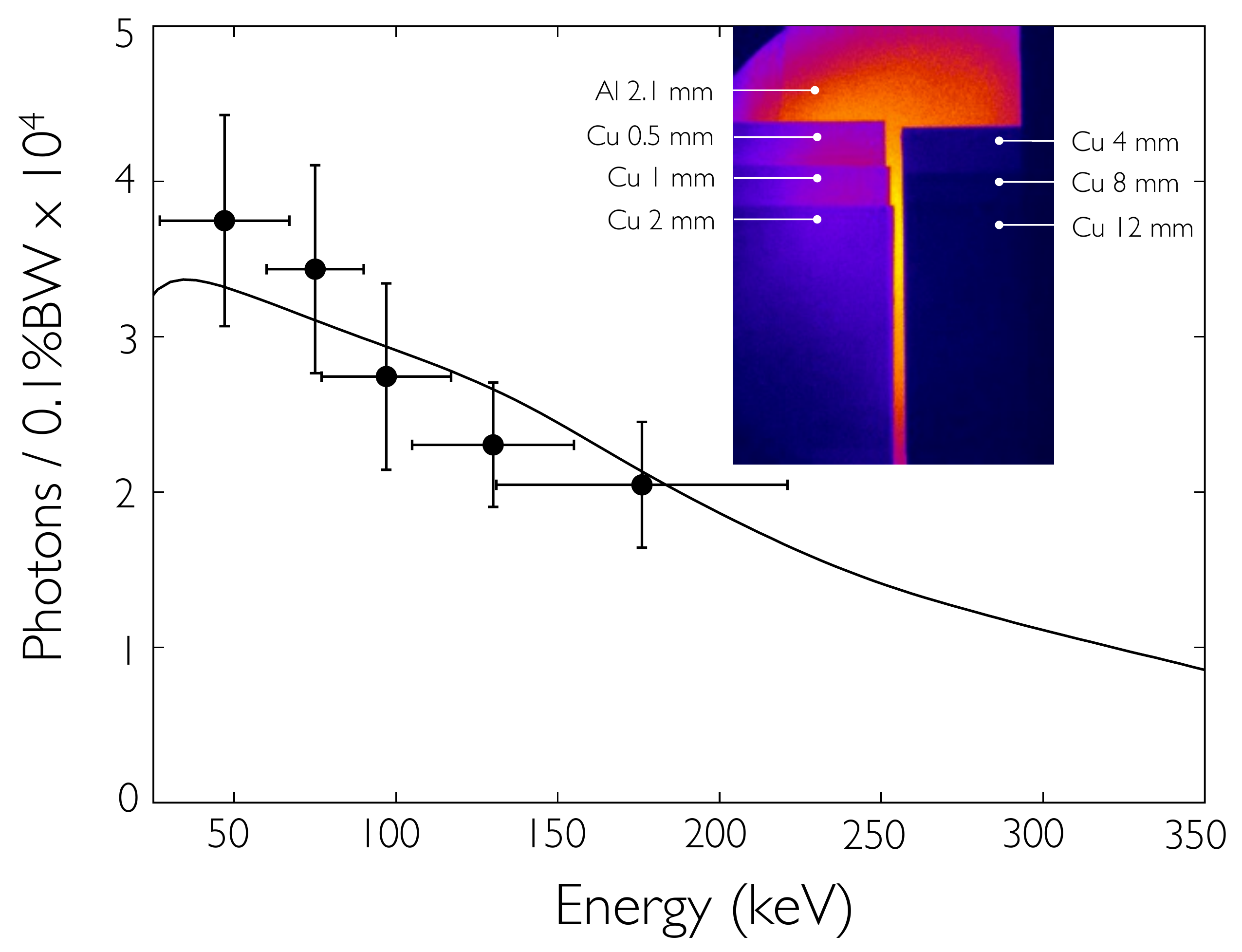}
\caption{\textbf{Spectrum obtained experimentally and numerical simulation.} The image shown in the inset is the beam profile obtained through a set of copper filters. It corresponds to an accumulation over 20 shots. The spectrum is therefore averaged over these 20 shots. The energy points of the spectrum correspond to the mean energy of the distributions $f_k$, and the horizontal error bars correspond to the FWHM width of these distributions (see Methods). The solid line corresponds to the calculated spectrum.}
\label{fig3}
\end{figure}

\begin{figure}[t]
\includegraphics[width=\linewidth]{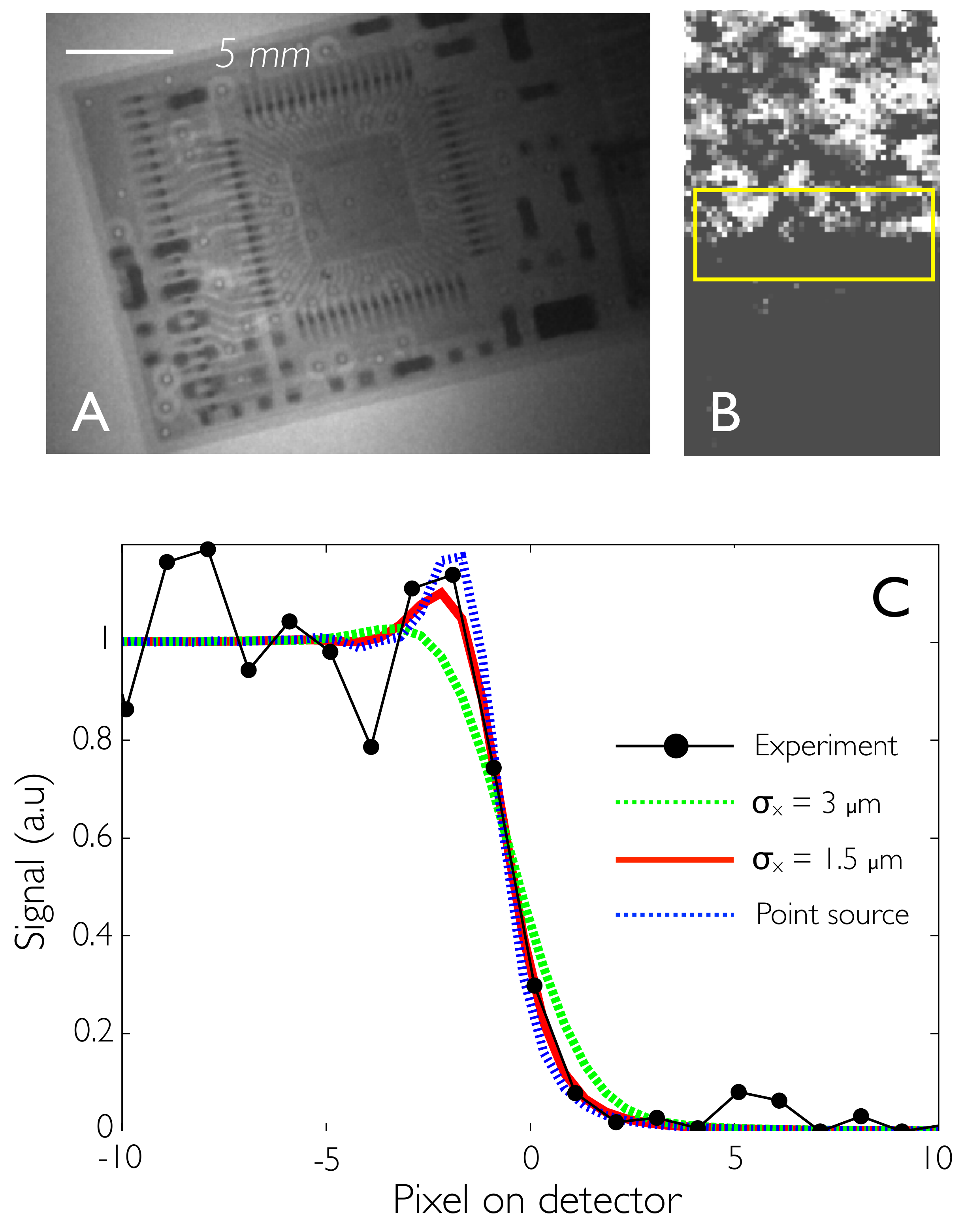}
\caption{\textbf{Radiography and source size measurement.} {\bf (A)} Radiograph of a USB flash drive with a magnification of  $\times1.2$. {\bf (B)} Knife edge image (knife edge is a cleaved InSb crystal). The detector is an imaging plate for the two images. {\bf (C)} Line-out of the knife edge image integrated over the region indicated in {\bf (B)}. Numerical fits are obtained by taking into account the Fresnel diffraction and considering the spectrum presented in Fig. \ref{fig3}.}
\label{fig4}
\end{figure}

The foil position $z$ along axis $\hat{z}$ is a crucial parameter in maximizing the Compton backscattering emission. It determines parameter $a_0$ of the reflected laser, which, together with $\gamma$, determines the radiation properties \cite{Catravas,Hartemann2}. Assuming a parallel electron beam and a head-on collision, the Compton backscattering radiation is emitted on axis at the fundamental wavelength $\lambda_{X}=(1+a_0^2/2)\lambda_L/4\gamma^2$ and at its harmonics if $a_0>1$. The amount of emitted energy scales as $a_0^2$ and $\gamma^2$.
Figure \ref{fig2} shows X-ray beams observed on a phosphor screen and the X-ray signal integrated over the angles as a function of position $z$ of the foil (here we used a 1 mm glass foil). After each shot, the foil is moved to provide undamaged surface, and foil position was measured using shadowgraphy. We define $z=0$ to be the position of maximum signal.
The behavior of the observed signal with $z$ is explained as follows. When the foil is moved outside the gas ($z>0$), $a_0$ at the position of collision rapidly decreases because the laser pulse, propagating in vacuum, diverges. In that case, the Compton backscattering flux becomes weak and the residual emission observed is due to the Bremsstrahlung radiation (this does not depend on $z$ and can be reduced below the detection threshold when the 1 mm glass is replaced by a 300 $\mu$m CH foil). When the foil is moved towards the inside of the gas jet ($z<0$), the properties of the electron beam are degraded, and the acceleration length is reduced to the point where injection does not occur. The X-ray signal diminishes as $z$ increases, and vanishes when no electrons are accelerated.
For $z=0$, the X-ray signal is at a maximum because the electrons have attained a sufficient energy and charge, and the strength parameter $a_0$ of the laser field, self-focused, is sufficiently high. In our experimental conditions, $z=0$ corresponds to the region of the exit gradient of the gas jet. At this optimum position, the X-ray beam divergence, which is governed by the divergence of the electron beam at the collision position, is 18 mrad (full-width-half-maximum, FWHM).
The spectrum of Compton backscattering depends on both the electron energy $\gamma$ and the strength parameter $a_0$.  In this experiment, the electron distribution was broadband and so is the X-ray spectrum. The X-ray energy distribution was measured using a set of copper filters with thicknesses from 500 $\mu$m to 1.2 cm, which covers the energy range from 50 keV to 200 keV. The image of the beam transmitted through the filters and the resulting spectrum are shown in Fig. \ref{fig3}. The X-ray emission extends up to a few hundred keV and the total photon number is $\sim1\times10^8$. To model the X-ray spectrum we used a test particle simulation that calculates the electron trajectories in the laser pulse and the resulting radiation (see Methods). A good agreement is found regarding both the energy distribution and the photon number, considering, as input parameters, the measured electron distribution, a charge of $120$ pC, and a $15$ fs (FWHM) laser pulse with $a_0=1.2$.

Such energetic radiation can be used to image the interior of a thick or dense object. As an example, Fig. \ref{fig4}A presents a radiograph of a USB flash drive obtained with the Compton source with a $\times1.2$ magnification. We also measured the X-ray transverse source size $\sigma_X$, as this determines the achievable resolution for large-magnification radiography applications, and the potential of the source for phase contrast imaging \cite{Wilkins}. For Compton backscattering, the X-ray source size is essentially equal to the transverse size of the electron bunch at the position of collision, which can be less than 1 $\mu$m. At the maximum emission position $z=0$, $\sigma_X$ was measured, in a single shot, using a knife edge technique. The edge was a cleaved InSb crystal located 20 cm from the source. The shadow of the X-ray beam formed by the sharp edge was recorded using an imaging plate placed 5 m from the source. Figures \ref{fig4}B-C present the image and corresponding line-out. As has been done previously for a laser broadband X-ray source \cite{ShahPRE2006}, we plotted the calculated Fresnel edge diffraction for the spectrum presented in Fig. \ref{fig3} \cite{BornWolf}. We obtain $\sigma_X < 3$ $\mu$m (FWHM) with a best fit for $\sigma_X =1.5$ $\mu$m (FWHM). The duration of the X-ray flash, essentially determined by the duration of the electron bunch, is expected to be a few tens of femtoseconds.

All the features of the measured radiation meet the properties of Compton scattering and no other known mechanisms can explain the results. As discussed earlier, the presence of the foil does not significantly modify the features of the electron beam and cannot therefore enhance the betatron and Bremsstrahlung radiations. Furthermore, betatron radiation in the 100 keV range, produced by 100 MeV electrons, would have much larger divergence (superior than a radian) and source size than the observed X-ray beam.  Bremsstrahlung can also be rejected, given the strong variation with the foil position and independence of the signal regarding foil material and thickness.

In conclusion, we have demonstrated the possibility of combining high energy X-ray photons, micrometre source size and a duration of tens of femtoseconds in a unique source. The method relies on a simple and original scheme of Compton backscattering in a laser-plasma accelerator. Based on the use of a single laser pulse, this scheme guarantees a systematic overlap between the laser pulse and electron bunch, and efficient use of the laser energy. In this experiment, X-ray beams are generated in a broadband energy range, extending up to a few hundred keV. The X-ray radiation is collimated within a 18 mrad (FWHM) beam with a source size of less than 3 $\mu$m (FWHM), and the total number of X-ray photons is estimated to be $\sim 10^8$. The peak brightness is of the order of $1\times10^{21}$ ph s$^{-1}$ mm$^{-2}$ mrad$^{-2}$ per 0.1$\%$ bandwidth at 100 keV. The high brightness with respect to conventional Compton sources is a result of the micrometre-scale source size and the femtosecond duration, features inherent to this all-optical scheme. Although the peak brightness is slightly inferior to a betatron source, the Compton source offers much higher photon energy at a given electron energy.
Future developments will focus on the production of nearly monochromatic and tunable radiation using the  monoenergetic and few-femtosecond electron bunches that can now be produced in a laser-plasma accelerator \cite{Faure1,Rechatin,Olle,Faure2}. Finally, the efficiency of the mechanism opens up the possibility of producing a compact femtosecond X-ray source using smaller high-repetition-rate laser systems with automated scanning of the plasma mirror substrate. 

\bigskip
\noindent
{\textbf{Methods}}

{\small

\noindent
\textbf{Laser system and target.} 
The experiment was conducted at Laboratoire d'Optique Appliqu\'ee with the ``Salle Jaune'' Ti:Sa laser system, which delivers 1 J / 35 fs (FWHM) pulses at a central wavelength of 810 nm and with a linear polarization. The laser was focused into a 3 mm supersonic helium gas jet with a 70-cm-focal-length off-axis parabola, to a focal spot size of 17 $\mu$m (FWHM). The laser pulse normalized amplitude in vacuum was estimated to be $a_0\simeq 1.5$ ($a_0=0.855\times10^{-9}\lambda_L[\textrm{$\mu$m}] \sqrt{I[\textrm{W/cm$^2$}]}$, where $\lambda_L$ is the laser wavelength and $I$ the intensity). The density profile of the gas jet along the laser propagation axis consists of a plateau of 2.1 mm and 600 microns gradients on both sides. The electron density in the plateau was $n_e \simeq 1.1\times 10^{19}$ cm$^{-3}$. The position of the laser focal plan was set in the first half of the gas jet to optimize the properties of the produced electron beam (charge and energy). 

\noindent
\textbf{Electron and X-ray beam diagnostics.} The electron spectrometer consisted of a permanent bending magnet (1.1~T over 10~cm) which deflects the electrons depending on their energy, and a Lanex phosphor screen to convert a fraction of the electron energy into 546 nm light imaged by a 16-bit visible charge-coupled device (CCD) camera. The minimum energy that can be measured was 50 MeV. In Fig. \ref{fig2}, the X-ray radiation was measured in a single shot using a Gd$_2$O$_2$S$:$Tb phosphor screen (Roper scientific) imaged with a 16-bit visible CCD camera. For Figs. \ref{fig3} and \ref{fig4}, we used a Fuji BAS-SR calibrated imaging plate (the images in Figs. \ref{fig3} and \ref{fig4}A correspond to an accumulation over 20 shots).

\noindent
\textbf{Spectrum analysis.} We define $T_k(\hbar\omega)$, the transmission of each filter indexed by $k=0,1,2,...$ (from thiner to the thicker), and $R(\hbar\omega)$ the response of the imaging plate (conversion from photon number to imaging plate counts). The signal transmitted by the filter $k$ was given by 
$
S_k = \int d(\hbar\omega)\: T_k(\hbar\omega)R(\hbar\omega) dN_\gamma/d(\hbar\omega),
$
where $dN_\gamma/d(\hbar\omega)$ is the photon number per eV and in one pixel. To obtain the X-ray spectrum we considered the signal difference between to successive filters, which is given by 
{\footnotesize
\begin{equation}
\nonumber
S_k-S_{k+1} = \int d(\hbar\omega)\: \frac{ [T_k(\hbar\omega)-T_{k+1}(\hbar\omega)]R(\hbar\omega)}{10^{-3}\hbar\omega} \left[ 10^{-3}\hbar\omega \frac{dN_\gamma}{d(\hbar\omega)} \right] .
\end{equation}}
We consider that the function $f_k=(T_k-T_{k+1})R/(10^{-3}\hbar\omega)$ is sufficiently narrow to assume that the number of photons per $0.1\%$ bandwidth $10^{-3}\hbar\omega \:dN_\gamma/d(\hbar\omega)$ is constant over the width of the $f_k$ distribution. Under this assumption, the above relation can be inverted to obtain the number of photons per $0.1\%$ bandwidth, given by $\left [10^{-3}\hbar\omega \: dN_\gamma / d(\hbar\omega) \right]_{|\hbar\omega_k} = (S_k-S_{k+1}) / \int d(\hbar\omega)\: f_k(\hbar\omega)$. The energy $\hbar\omega_k$ assigned to this photon number is the mean of the distribution $f_k$. The horizontal error bars correspond to the FWHM of the $f_k$ distributions.  

\noindent
\textbf{Numerical modeling.} We used a test particle simulation based on a Runge-Kutta algorithm. The code calculates the orbit of one electron oscillating in a counterpropagating laser pulse characterized by a normalized vector potential $\mathbf{a}=a_0\exp[-(2\ln2)(z+ct)^2/(c^2\tau^2)]\cos(\omega_it+k_iz)\mathbf{e}_x$, where $\tau$ is the FWHM pulse duration. Once the trajectory calculated, the radiated energy per unit solid angle and unit frequency is obtained by integrating the general formula for the radiation from a moving charge \cite{Jackson}. The result is then integrated over the angular distribution, converted into photons per $0.1\%$ bandwidth. This calculation is performed for all electron energies, and the Compton spectrum is finally obtained by integrating over the experimental electron spectrum, which is presented in Supplementary Fig. S1. The radiation damping is not considered because the energy of one electron is much larger than the X-ray energy it radiates.
}

\bigskip
\noindent
{

{\small

\bigskip
\noindent
{\textbf{Acknowledgments.}} The authors acknowledge the European Research Council through the PARIS ERC project (Contract No. 226424). The authors acknowledge LOA technical staff for experimental assistance.

\bigskip
\noindent
{\textbf{Authors contributions.}} K.T.P., S.C. and C.T. have conceived and realized the experiment, and contributed equally to this work. K.T.P., S.C., C.T. and V.M. analyzed the data. K.T.P., S.C., C.T., V.M., R.C. and A.R. wrote the paper. J.P.G. and A.T. operated the laser system. R.S. proposed the experiment. V.M., A.R. and S.S. supported the project.

}

\clearpage

\begin{figure*}[t]

\includegraphics[width=0.7\linewidth]{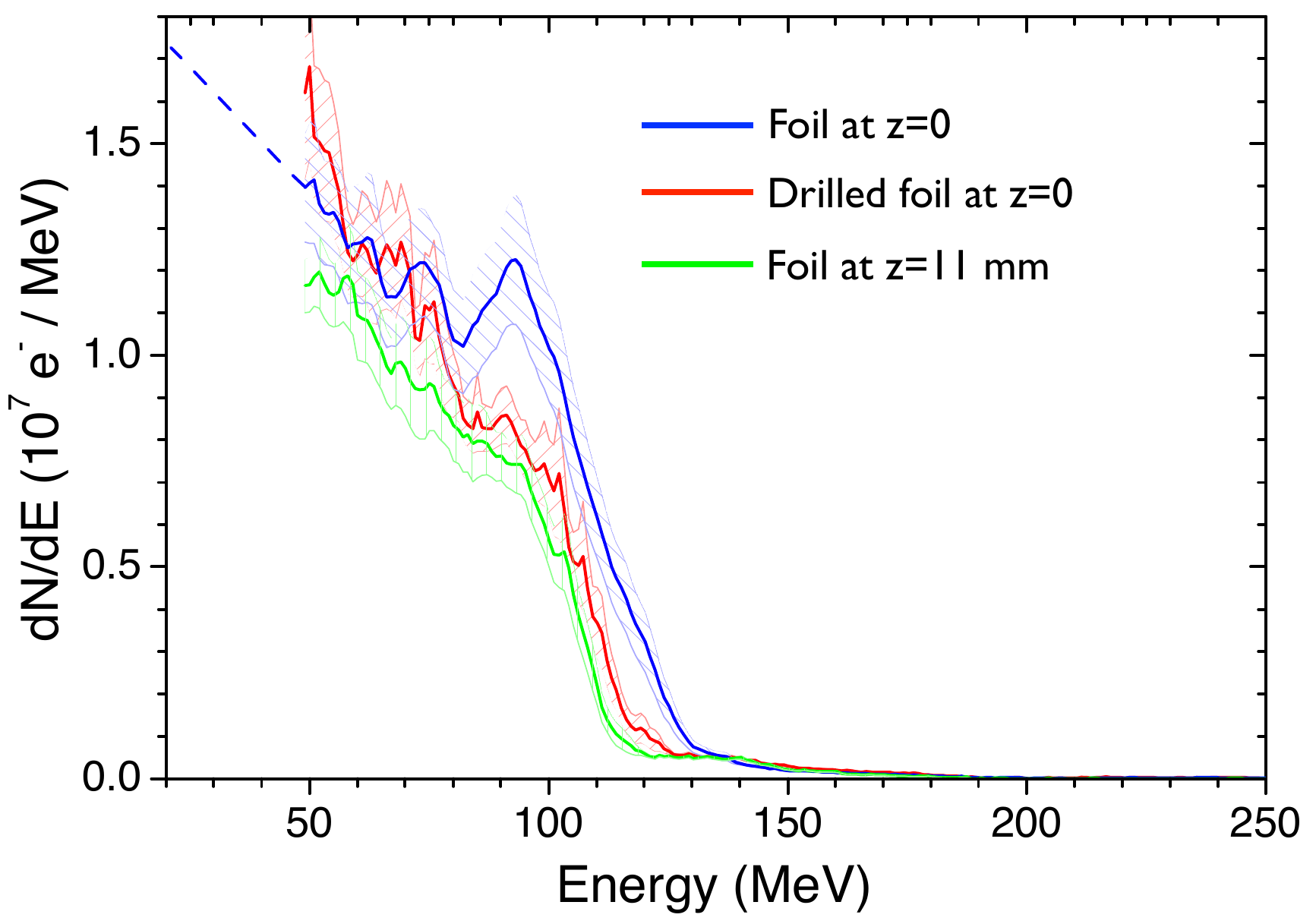}\\
\label{figS1}
\begin{flushleft}
{\normalsize \textbf{Supplementary Fig. S1.} Electron spectra experimentally measured for the foil in the gas at $z=0$, for the drilled foil at $z=0$ (drilled along the laser axis), and for the foil out of the jet $z = 11$ mm. Comparing these spectra demonstrate that the foil and the laser interaction with the foil can not significantly improve the electron beam. The presented spectra are averaged over 20 shots. The spectrum used for the numerical modeling of the Compton spectrum is linearly extrapolated for energies below 50 MeV.}
\end{flushleft}
\end{figure*}

\end{document}